\documentclass[aps,prl,preprint,superscriptaddress,amsmath,amssymb]{revtex4}
\usepackage{graphicx}
\usepackage{color}
\begin{document}
\title{Molecular symmetry effects in the ionization of CS$_2$ by intense, few-cycle laser pulses}
\author{Deepak Mathur}
\email{atmol1@tifr.res.in}
\affiliation{Tata Institute of Fundamental Research, 1 Homi Bhabha Road, Mumbai 400 005, India}
\affiliation{UM-DAE Centre for Excellence in Basic Sciences, University of Mumbai - Kalina Campus, Mumbai 400 098, India}
\author{Aditya K. Dharmadhikari} 
\affiliation{Tata Institute of Fundamental Research, 1 Homi Bhabha Road, Mumbai 400 005, India}
\author{Firoz A. Rajgara}
\affiliation{Tata Institute of Fundamental Research, 1 Homi Bhabha Road, Mumbai 400 005, India}
\author{Jayashree A. Dharmadhikari}
\affiliation{UM-DAE Centre for Excellence in Basic Sciences, University of Mumbai - Kalina Campus, Mumbai 400 098, India}

\begin{abstract}
Few-cycle pulses of intense 800 nm light are used to probe ionization and dissociation of carbon disulfide in the intensity and temporal regime where rescattering is expected to dominate the laser-molecule interaction. The wavepacket of the rescattered electron destructively interferes with the anti-bonding $\pi$-orbital of CS$_2^+$ such that rescattering is effectively ``switched off". Direct signature of enhanced ionization being ``switched off" in the ultrashort domain is also obtained. Consequently, dissociation becomes an almost non-existent channel when few-cycle pulses are used, with only long-lived singly-, doubly-, and triply-charged molecular ions dominating the mass spectrum. Few-cycle optical pulses help reveal that quantum-mechanically determined molecular symmetry contributes to strong field molecular ionization.
\end{abstract}
\maketitle

Studies of the interaction of intense laser fields with molecules continue to provide a steady stream of discoveries of unexpected, sometime counter-intuitive, phenomena and processes that constantly invigorate strong-field science (for a recent compilation of cogent reviews, see \cite{PUILS}, and references therein). In such studies, the magnitude of the optical field matches the intra-molecular Coulombic field, and the overall laser-molecule interaction is dominated by single and multiple ionization, leading to the breaking of one or several bonds. Most experimental probes of how molecules behave in strong fields have focused on measurement of ion yields, usually with infrared laser pulses of durations ranging from a few tens to a few hundred femtoseconds, and it seems established that enhanced ionization (EI), spatial alignment, and rescattering ionization are the main drivers of the dynamics \cite{PUILS}. On the other hand, very recent work \cite{wu} shows that when few-cycle pulses are used, the dynamics become significantly different. Dynamic alignment of molecules like O$_2$, N$_2$ does not occurs, with the duration of the optical field being too short for the polarization-induced torque to act on the molecular axis \cite{tong}.  Similarly, as the few-cycle dynamics proceed essentially at equilibrium bond lengths, EI is effectively ``switched off"; nuclei within the irradiated molecule do not have enough time to move to the critical distance \cite{tong} at which the propensity for ionization is enhanced. The dynamics in the few-cycle regime are, therefore, considerably simplified, dominated by only rescattering, wherein the electron that is produced by optical field ionization oscillates in the optical field so as to collide with the parent molecular ion, inducing further ionization. Few-cycle pulses, therefore, offer the prospect of disentangling the effect that different processes make to strong-field molecular dynamics. Moreover, as experiments on H$_2$ have shown, control on the dynamics may also be exercised by tuning the intensity and duration of the ultrashort optical field \cite{alnaser2004}. But what of the quantal structure of the molecule itself? Does it have any role to play? Does the few-cycle, strong-field regime take cognizance of molecular symmetry, a consequence of the quantum-mechanically determined electronic structure? We explore these questions in experiments on irradiation of CS$_2$ molecules by intense, four-cycle pulses that we report in this Letter. 

Before rationalizing why we choose to probe CS$_2$ in order to assess the possible role of quantal effects in the strong field regime, we make the following observations. A plethora of data on the behavior of atoms in strong-fields has provided confirmation that the multiphoton ionization rate, and the resulting yield of ions, depends only on one atomic property: the first ionization energy (IE). Even in tunnel ionization, wherein the optical field distorts the atom's radial potential function, enabling one or more valence electrons to tunnel into the continuum, the resulting electron energy distribution is well accounted for by the oft-used ADK (Ammosov-Delone-Krainov) theory~\cite{ADK} in which the only atomic parameter of concern is only the lowest IE; the quantal nature of the orbitals themselves does not enter into reckoning. Correspondingly, nearly identical ionization rates were measured for Ar and N$_2$, where the ratio IE(Ar)/IE(N$_2$) is 1.01 \cite{talebpour}. Unexpectedly, the ionization rate for the O$_2$ molecule was measured to a factor of ten lower than that for the Xe atom in spite of the ratio IE(Xe)/IE(O$_2$) being even closer to unity (1.005). Rationalizations proferred in the orignal reports \cite{talebpour} invoked either multielectron effects or the nuclear motion within molecules. Multielectron effects failed to succeed by way of explanation as the resulting ADK ionization yields \cite{ADK} only matched measured ones at laser intensities beyond the saturation intensity \cite{grasbon}. It has been noted \cite{grasbon} that calculations \cite{saenz} show that molecular vibrations also fail to quantitatively rationalize differences between molecular ionization rates and those of the companion atoms. An entirely different insight has emerged from intense field $S$-matrix calculations \cite{faisal} that predict suppression of ionization in homonuclear molecules with an antibonding valence orbital, like the outermost $\pi_g$ orbital in O$_2$, but not in the case of molecules with a bonding valence orbital, like the $\sigma_g$ orbital in N$_2$. The shape of the former orbitals results in destructive interference by the two nuclei of subwaves of the ejected electron. Electron spectroscopy of diatomics has recently offered compelling vindication of this projected scenario \cite{n2o2}.

We focus here on the linear triatomic, CS$_2$, in order to probe the possible role of symmetry effects in strong-field molecular dynamics in the few-cycle domain. Neutral and ionized CS$_2$ molecules are known to be important intermediaries in chemical transformation processes in cold interstellar plasmas, cometary environments and in planetary and interstellar atmospheres \cite{vardya,cosmovici}. Also, CS$_2$ is an efficient ionizing agent in charge-exchange organic mass spectrometry \cite{mercer}. From the perspective of this work, we note that the ground electronic state of CS$_2$ has the electronic configuration ${\rm (Core)}^{22}~(5\sigma_g)^2(4\sigma_u)^2(6\sigma_g)^2(5\sigma_u)^2(2\pi_u)^4(2\pi_g)^4$, which yields overall symmetry, $^1\Sigma_g^{~+}$. The outermost $2\pi_g$ orbital is mostly built up of 3p orbitals of the S-atoms; as the equilibrium bond length is large ($\sim$1.6\AA), there is little $\pi$-overlap between the two peripheral atoms and, consequently, the anti-bonding character dominates single ionization that occurs upon removal of an electron from the antibonding orbital; double and triple ionization occurs upon removal of successive electrons from the same antibonding orbital. Each such removal effectively enhances the electronic charge density in the internuclear region of the molecule, resulting in long-lived doubly- and triply-charged molecular ions, CS$_2^{2+}$ and CS$_2^{3+}$ whose lifetimes have previously been measured to be of the order of seconds \cite{aarhus}. 

Previous work on few-cycle ionization dynamics appears to have been conducted using the hollow-fibre pulse compression technique \cite{nisoli}. Recently, few-cycle pulses have also been generated using filamentation in gas-filled tubes \cite{gas}. We used here 0.5 mJ, 50 fs laser pulses centered at 800 nm (1 kHz repetition rate) from a Ti-sapphire amplifier. After passing through an aperture the laser beam was focused with a 1.0 m focal length metal coated spherical mirror on to a 1.5 m long tube containing 1.2 atm of Ar. The central part of the resulting broadband light was compressed using a set of chirped dielectric mirrors (CDM) to produce 15 fs pulse (0.3 mJ energy). These pulses were then passed through another aperture and focused on to a 1 m long second tube (filled with 0.9 atm of Ar). The broadband light was again compressed using a second pair of CDM to yield 11 fs (four-cycle) pulses with energy of 0.25 mJ. The compressed pulses were characterized using spectral phase interferometry for direct electric field reconstruction and were directed through a 300 $\mu$m fused silica window into an ultrahigh vacuum (UHV) chamber in which the laser-CS$_2$ interaction occured at pressures low enough ($\sim$10$^{-9}$ Torr) to obviate the need to consider space charge saturation effects. The four-cycle pulses were precompensated for the chirp and were focused within the UHV chamber by a spherical mirror ($f$=5 cm). Ion analysis was by conventional time-of-flight (TOF) methods. 
\begin{figure}
\includegraphics[width=6cm]{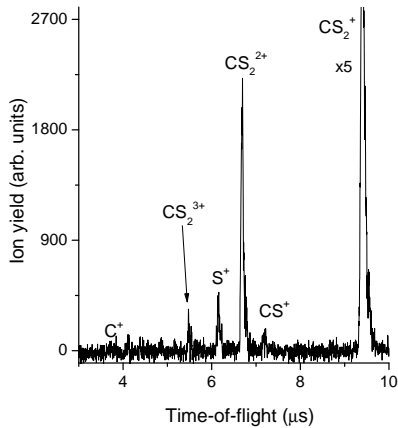}
\caption{Typical time-of-flight spectrum of CS$_2$ obtained at an intensity of 5$\times$10$^{15}$ W cm$^{-2}$ using four-cycle (11 fs) pulses of 800 nm light.}
\end{figure}
A typical TOF spectrum that we obtain at intensity 5$\times$10$^{15}$ W cm$^{-2}$ is shown in Fig. 1. The two striking features of this spectrum are (i) the dominance of peaks corresponding to different charge states of CS$_2$, and (ii) the absence of fragment ion peaks that dominate spectra obtained with longer-duration pulses, an example of which is presented in Fig. 2a for 40 fs pulses of 800 nm light at intensity 6.7$\times$10$^{15}$ W cm$^{-2}$. Very similar spectra were obtained when we used 100 fs pulses of this intensity. The dissociative ionization pattern is clearly much richer with the longer pulse, with a gamut of atomic fragments being produced up to charge state 4+. Ions like S$^+$ and S$^{2+}$ are produced with substantial kinetic energy ($\sim$4 eV), as exemplified by the forward-backward peak splitting of the TOF peaks corresponding to these fragment ions, the precursors being excited states of highly-charged CS$_2^{q+}$ ($q>$1) that Coulomb-explode. Long-lived CS$_2$ dications and trications are observed, but with much lower overall yield compared to the fragment ions. The stark difference between spectra in Figs. 1 and 2a is obvious and lies in the dramatic suppression of the fragmentation channels when few-cycle pulses are used. How do we rationalize this difference?
\begin{figure}
\includegraphics[width=6cm]{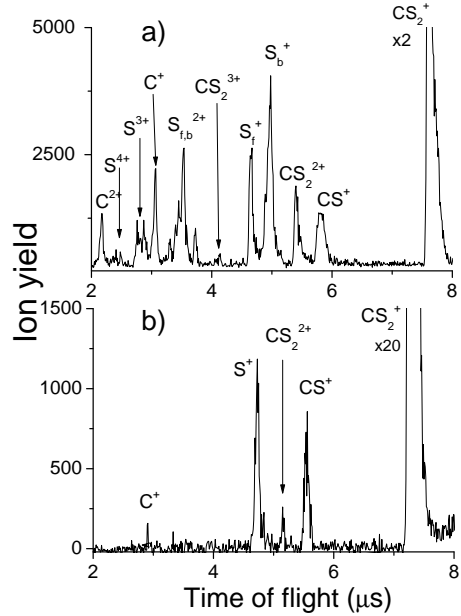}
\caption{a) Time-of-flight spectrum of CS$_2$ obtained at intensity 6$\times$10$^{15}$ W cm$^{-2}$ using 40 fs pulses of 800 nm light. Very similar spectra were measured using 100 fs pulses. Note the preponderence of fragment ions. S$_f^+$, S$_b^+$ denote singly-charged S-ions the Coulomb-explode in a direction towards and away from the detector at the end of our TOF spectrometer. b) Corresponding spectrum using broadband light (bandwidth = 500-900 nm).}
\end{figure}
Is the spectral bandwidth of the incident light a factor that needs to be taken cognizance of? We made TOF measurements on CS$_2$ using broadband light, with bandwidth $\sim$500-900 nm (Fig. 2b). This supercontinuum was generated by irradiating 1 cm long BK-7 glass with intense 800 nm light; the methodology adopted for such white light experiments on molecules, including estimation of the white light intensity, has been described elsewhere recently \cite{methaneJPC}. The resulting TOF spectrum also shows dissociation occuring, with prominent fragments S$^+$ and CS$^+$, whose yields relative to, say, CS$_2^{2+}$ are consistent with those obtained in data obtained using single-color light pules of 40, 100 fs duration (Fig. 2a). We are led to conclude that the broadband nature of our four-cycle pulses is not an important factor in the observed suppression of fragmentation channels in CS$_2$. 

One major clue to the role that the quantal description of the irradiated molecule might be of significance emanates from our observation of fragments like S$^+$ and CS$^+$ when we use longer pulses. S$^+$ and CS$^+$ cannot be produced by direct ionization of CS$_2$ since Franck-Condon factors preclude vertical access to the dissociation continua of the $X$, $A$ and $B$ electronic states of CS$_2^+$. We base this conjecture on the absence of any peak in the CS$_2$ photoelectron spectrum \cite{Brundle} near the appearance thresholds for S$^+$ and CS$^+$ fragments (14.81 and 15.78 eV, respectively), and also from results of photoion-photoelectron coincidence measurements \cite{Brehm}.  The next ionic state, $C$, lies above the dissociation limits S$^+$ + CS and S + CS$^+$ and, hence, fully predissociates. This has also been established in photoion-photoelectron coincidence measurements \cite{Brehm}. It is the lengthening of the C-S bond in the EI process that properly and satisfactorily accounts for the prominent yield of S$^+$ and CS$^+$ fragments when we make measurements with long pulses: population of excited electronic states of CS$_2^+$ becomes likely and these then are precursors of the fragments in question. The disappearance of these fragments from our four-cycle spectra is a clear and direct signature that the EI process is switched off in the ultrashort domain. This conjecture is also quantitatively supported by estimating timescales for dissociation of multiply-charged CS$_2$ precursors under the assumption that the di- and tri-cation potential energy surfaces are dominated by the Coulomb term. Take the peripheral S-atoms to have reduced mass $M$ at an initial separation of 2$\beta a_o$. Following field ionization to charge states $Z_1$ and $Z_2$, the time, $\tau$, taken for the products of Coulomb explosion to develop a separation of $x$ is \cite{time}
\begin{equation}
\tau \sim  \frac{\rm \lambda_{c}}{\rm 2\pi c\alpha_{2}} 
{\Biggl[{\frac {\rm \beta^{3}M}{\rm 2m_{e}Z_{1}Z_{2}}} \Biggr]}^{1/2} 
{\Biggl[ {\frac{\rm x(1-{2\beta a_0}/x)^{1/2}}{\rm 2\beta a_0}} +
{\frac {\rm 1}{\rm 2}}ln{\biggl[ {\frac {\rm 1+
{(1-{2\beta a_0}/x)}^{1/2}}{\rm 1-
{(1-{2\beta a_0}/x)}^{1/2} }} \biggr]} \Biggr]},
\end{equation}
where $\alpha$ is the fine-structure constant, $m_e$ is the electron mass, $\lambda_c$ is the Compton wavelength while $a_0$ and $c$ are the Bohr radius and speed of light, respectively. The time dependences of dissociation of various charge states can be readily estimated. For instance, following Coulomb explosion of CS$_2^{2+}$, it takes as long as $\sim$40 fs for S$^+$-S$^+$ ion-pairs to seperate to 10{\AA}, far too long for a four-cycle pulse. 

But what of recscattering dynamics? This occurs on ultrafast timescales but, as we discuss in the following, is another area where quantal considerations cannot be ignored. We note that in double ionization of H$_2$ \cite{alnaser2004} induced by intense pulses shorter than $\sim$15 fs, the first return recollision essentially dominates the rescattering dynamics whereas for longer pulses, the third return recollision could assume importance. Experiments on methane \cite{wu} also highlight the importance of the first return recollision when 8 fs pulses are used: no doubly charge ions were observed with 8 fs pulses while such ions began to appear when longer pulses were used. In the present work on CS$_2$, however, doubly- and triply-charged parent ions dominate the four-cycle spectrum, ostensibly at the expense of molecular fragmentation channels. This is a signature of rescattering also being ``switched off" in our experiments not because of temporal constraints but those imposed by the quantum-mechanical nature of CS$_2$'s outermost $2\pi_g$ orbital. The wavepacket of the returning electron interferes destructively with the spatial extent of this orbital, leading to effective cancellation of the rescattering process. The returning electron's energy is no longer available for electronic excitation to CS$_2^{+*}$ states that are quantally allowed to dissociate into S$^+$+CS$^+$. 

There is another, totally different scenario in which strong field effects occur for ultrashort durations {\em and} without the possiblity of rescattering: in large impact-parameter collisions of highly charged ions with molecules \cite{benzene}. We have studied 100 MeV collisions of CS$_2$ with Si$^{8+}$ ions in a tandem accelerator and measured ion spectra for impact parameters larger than 3 {\AA}. The methodology has been described elsewhere \cite{benzene} but we note here that the chosen ion charge state and impact parameter ensure that the CS$_2$ molecule experiences fields of comparable magnitude to those in our four-cycle laser experiments. The collision energy ensures that the interaction time is ultrashort, of the order of 0.7 fs. The resulting spectrum (Fig. 3) is remarkable in its similarity to that obtained with four-cycle pulses (Fig. 1) and confirm that rescattering plays little or no role in the strong field dynamics of CS$_2$ when four-cycle optical pulses are used.
\begin{figure}
\includegraphics[width=6cm]{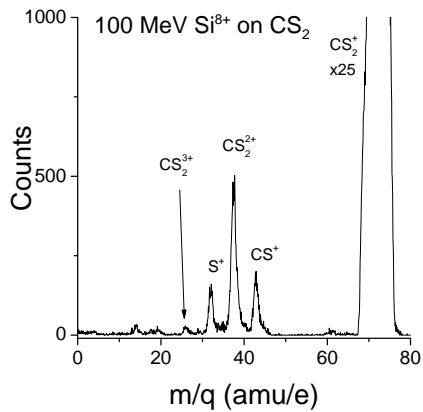}
\caption{Mass spectrum of CS$_2$ in collision with 100 MeV Si$^{8+}$ ions.}
\end{figure}
Our CS$_2$ results offer unambiguous vindication of the strong field $S$-matrix theory developed for diatomic molecules \cite{faisal}: we predict that destructive interference brought about by the antibonding $\pi_g$ orbital of CS$_2^+$ will result in suppression of the plateau region in electron spectra. Such suppression would not be observed in the case of the bonding $\sigma_g$ orbitals. We note that electron density profles of the N$_2$ and O$_2$ HOMOs have been mimicked by Alnaser {\it et al.} \cite{alnaser2004} using intense 8 fs pulses. More recently, Coulomb explosion studies of N$_2$ with 10 fs pulses have confirmed that there is no significant stretching of the N-N bond on such timescales \cite{baldit}. We believe that intense few-cycle pulses offer a new route to Coulomb explosion imaging of neutral molecules in their equilibrium (un-stretched) geometry.

We are grateful to the Department of Science and Technology for partial financial support of this work.

\end{document}